\documentclass[aps,prc,twocolumn,groupedaddress,showpacs]{revtex4-1}

\usepackage{graphicx}
\usepackage{dcolumn}
\usepackage{bm}
\usepackage{amssymb,amsmath}
\usepackage{enumerate}

\begin{document}

\title{Precise Half-Life Measurement of the Superallowed $\beta^{+}$ Emitter $^{46}$V}

\author{H.I. Park}
\email[]{hpark@comp.tamu.edu}
\author{J.C. Hardy}
\email[]{hardy@comp.tamu.edu}
\author{V.E. Iacob}
\author{L. Chen}
\author{J. Goodwin}
\author{N. Nica}
\author{E. Simmons}
\author{L. Trache}
\author{R.E. Tribble}

\affiliation{Cyclotron Institute, Texas A\&M University, College Station, Texas 77845-3366, USA}

\date{\today}

\begin{abstract}
The half-life of $^{46}$V has been measured to be 422.66(6) ms, which is a factor of two more precise than the best previous measurement. Our result is also consistent with the previous measurements, with no repeat of the disagreement recently encountered with $Q_{EC}$ values measured for the same transition.  The $\mathcal{F}t$ value for the $^{46}$V superallowed transition, incorporating all world data, is determined to be 3074.1(26) s, a result consistent with the average $\overline{\mathcal{F}t}$ value of 3072.08(79) s established from the 13 best-known superallowed transitions.  
\end{abstract}


\maketitle

\section{Introduction}
At present, the most stringent test of the unitarity of the Cabibbo-Kobayashi-Maskawa (CKM) matrix depends on results from superallowed $0^{+}$$\rightarrow$\,$0^{+}$ nuclear $\beta$ decay. Precise measurements of $ft$ values for these superallowed transitions provide a direct determination of the weak vector coupling constant $G_V$ within an uncertainty of $\pm$0.013$\%$ and lead to the most precise value of $V_{ud}$, the up-down quark-mixing element of the CKM matrix, within $\pm$0.023$\%$~\cite{to10}. Incorporating this result, the sum of squares of the top-row elements of the CKM matrix satisfies the unitarity condition at the level of $\pm$0.06$\%$~\cite{to10}. This remarkable agreement with the standard model constrains the scope of any new physics possible beyond the standard model, and motivates the quest for still higher experimental precision to make the unitarity test even more definitive.

The $ft$ value that characterizes a $\beta$-transition depends on three measured quantities: the total transition energy $Q_{EC}$, the half-life $t_{1/2}$ of the parent state, and the branching ratio $R$ for the particular transition of interest.  The statistical rate function $f$ depends on the fifth power of the $Q_{EC}$ value, while the partial half-life $t$ is given by the half-life divided by the branching ratio.  Thus, the fractional uncertainty required for $Q_{EC}$-value measurements must be five times smaller than that for half-life and branching-ratio measurements to have the same impact on the fractional uncertainty for the $ft$ value.  Within the past decade, the advent of on-line Penning traps has made a significant improvement possible in $Q_{EC}$ measurements, which has had an important impact on world data for superallowed beta decay.

The decay of $^{46}$V was the first of the nine best-known superallowed transitions to have its $Q_{EC}$ value measured (in 2005) with an on-line Penning trap~\cite{sa05}. The result differed significantly from the previously accepted result -- a longstanding reaction-based $Q_{EC}$ value published in 1977 \cite{vo77} -- and shifted the $^{46}$V $ft$ value two standard deviations out of agreement with other well-known superallowed transitions.  This apparent deviation from the conserved vector current (CVC) expectation raised several concerns, among them the possibility of systematic differences between reaction and Penning-trap measurements of $Q$ values. As a result, another independent Penning-trap measurement was performed a year later, in which the $Q_{EC}$ value for $^{46}$V was determined again, along with the $Q_{EC}$ values for two other superallowed beta emitters, $^{26}$Al$^{m}$ and $^{42}$Sc~\cite{er06}.  The second measurement for $^{46}$V confirmed the first Penning-trap result.  At the same time though, the $Q_{EC}$-value results for $^{26}$Al$^{m}$ and $^{42}$Sc were consistent with previous reaction-based values, thus demonstrating that there were no widespread systematic effects afflicting all reaction-based measurements.  Evidently, the problem with the $^{46}$V $Q_{EC}$ value arose from some flaw specific to one particular set of reaction measurements \cite{vo77}.  This conclusion was further strengthened by a subsequent $Q$-value measurement in 2009 \cite{fa09}, which used the same ($^3$He, $t$) reaction and some of the same equipment as had been used 32 years earlier in the flawed measurement.  It obtained a $Q_{EC}$ value for $^{46}$V that was entirely consistent with the two Penning trap results.  Finally, even more recently, another Penning-trap measurement of the $^{46}$V $Q_{EC}$ value has been published \cite{Er11} with a 100-eV uncertainty, a factor of three smaller than any previous measurement.  It agrees with the previous Penning-trap measurements and with the recent ($^3$He, $t$) result.

In parallel with these experimental studies, the theory used in the analysis of the $^{46}$V results was carefully reexamined as well.  In order to extract $G_V$ from a measured $ft$ value, radiative and isospin-symmetry-breaking corrections must both be applied to convert the measured $ft$ value into a corrected $\cal F$$t$ value \cite{ha09}.   In the case of $^{46}$V and the other $f_{7/2}$-shell nuclei, the original isospin-symmetry-breaking corrections had been derived with a shell model based on a closed $sd$ shell at $^{40}$Ca.  However the presence of low-lying $3/2^+$ and $1/2^+$ states in $^{45}$Ti and other nuclei in the region shows that core orbitals must play an important role.  Accordingly, the isospin-symmetry-breaking corrections were improved \cite{to08} by the inclusion of core orbitals in the model space used in the calculation.  The choice of which orbitals to include was guided by the measured spectroscopic factors for single-nucleon pickup reactions on the nuclei of interest.  This improvement had the largest effect on the $^{46}$V transition, but also had smaller effects on other superallowed transitions, especially on those in the $f_{7/2}$ shell.  These new corrections, which were incorporated into the most recent survey of world data \cite{ha09}, eliminated the $^{46}$V anomaly, restored the consistency with CVC and retained agreement with CKM matrix unitarity. 

All these experimental and theoretical studies have proceeded under the tacit assumption that the previously accepted half-life of $^{46}$V was completely correct.  Although there was no reason to suspect that this half-life result was in error, there had not been any reason to suspect that there was anything wrong with the now-discredited $Q_{EC}$ value either.  We have addressed this potential weakness and report here on a new measurement of the $^{46}$V half-life, which confirms the average of previous measurements but is a factor of two more precise than the best of them.  

\section{Experimental details}
\label{sec:exp}

After making concerted efforts over several years to develop a useful titanium beam from our ECR ion-source, we employed a metal foil, 94$\%$ enriched in $^{47}$Ti, as source material and successfully obtained a 32$A$ MeV $^{47}$Ti beam with up to 50 nA of current from the K500 superconducting cyclotron at Texas A$\&$M University.  Directing this beam into a hydrogen gas target, we produced $^{46}$V via the inverse-kinematics reaction $^1$H($^{47}$Ti, 2$n$)$^{46}$V. The hydrogen gas cell had 4-$\mu$m-thick Havar entrance and exit windows.  To increase the gas density and the yield of reaction products, it was operated at liquid-nitrogen temperature and at 2-atm pressure.  The fully-stripped reaction products exiting the target cell entered the Momentum Achromat Recoil Spectrometer (MARS) \cite{tr89}, where they were separated according to their charge-to-mass ratio $q/m$, with $^{46}$V being selected in the focal plane.

\begin{figure}[t]
\centering
\includegraphics[width=\columnwidth]{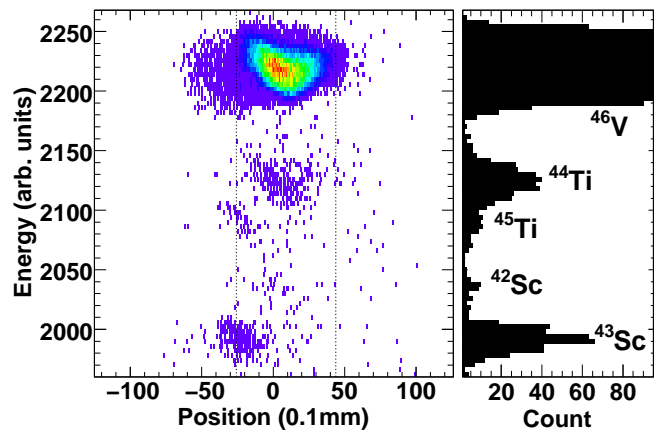}
\caption{\label{fig1} (Color online) Deposited energy versus position as recorded in the PSSD at the MARS focal plane.  This result was obtained after the spectrometer had been tuned for $^{46}$V. The dashed lines show the position of the extraction slits, 7 mm apart, which we used during these measurements.  Note that the tail to the left of the main $^{46}$V peak is an artifact caused by incomplete charge collection in the strip detector.  From such spectra recorded periodically during our experiment, we determined that the extracted $^{46}$V beam included a 0.12\% contribution from $^{42}$Sc.}
\end{figure}

Initially working with a low-current primary beam, we inserted a 1-mm-thick 16-strip position-sensitive silicon detector (PSSD) at the focal plane of MARS. This detector was used first for the identification of secondary reaction products, then for the control of the selection and focus of $^{46}$V in the center of the beam line. As shown in Fig.~\ref{fig1}, in addition to $^{46}$V, there were four reaction products, $^{42}$Sc ($t_{1/2}$\,=\,680.72 ms), $^{43}$Sc ($t_{1/2}$\,=\,3.891 h), $^{44}$Ti ($t_{1/2}$\,=\,60.0 y) and $^{45}$Ti ($t_{1/2}$\,=\,3.08 h), that appeared between the extraction slits and were thus weak contaminants in the extracted $^{46}$V beam.  The presence of $^{44,45}$Ti and $^{43}$Sc was not problematic since their half-lives are more than four orders of magnitude longer than 423-ms $^{46}$V.  Our only concern was $^{42}$Sc, another superallowed $\beta$-emitter with a rather similar half-life to that of $^{46}$V.  With the focal-plane acceptance slits of MARS set to a width of 7 mm, the total extracted beam contained 0.12$\%$ of $^{42}$Sc nuclei. The composition of the beam exiting MARS was checked on a daily basis during our half-life measurement: on each occasion we reinserted the PSSD at the MARS focal plane and acquired a spectrum equivalent to the one shown in Fig.~\ref{fig1}. There were no appreciable changes observed in the extracted beam composition at any time. 

We removed the PSSD and increased the primary beam intensity after the tune-up procedures were completed. The $^{46}$V beam from the extraction slits then exited the vacuum system through a 51-$\mu$m-thick Kapton window, passed through a 0.3-mm-thick BC404 scintillator, where the ions were counted, and then through a stack of aluminum degraders, finally stopping in the 76-$\mu$m-thick aluminized Mylar tape of a fast tape-transport system. A sample was collected in the tape for a fixed time -- either 0.3 s or 0.5 s -- after which the cyclotron beam was interrupted and the activity moved in 180 ms to the center of a 4$\pi$ proportional gas counter located at a well-shielded area about 90 cm away.  The length of the collection period was chosen to obtain an initial $\beta$-particle counting rate of between 5000 and 10000 particles/sec; it was altered when necessary to compensate for changes in the primary beam intensity.  The 4$\pi$ proportional gas counter and associated electronics were the same as those originally described in \cite{ko97} and again in the reports of our previous half-life measurements ({\it e.g.} Refs. \cite{ia06,ia10}); the gas counter was always operated in the plateau region.

Multiscaled signals from the gas counter were recorded for 10 s into two separate 500-channel time spectra, each corresponding to a different pre-set dominant dead-time. Having these two data streams allowed us to test that our dead-time corrected results were independent of the actual dead time of the circuit (see Sec.~\ref{ssec:tdec}). The time increment per channel for the multichannel scalars was externally controlled by a function generator accurate to 0.01 ppm and the two different non-extending dead times were continuously monitored on-line to an accuracy of $\pm$5 ns. The collect/move/count cycle timing was continuously monitored on-line as well. The cycles were repeated until the desired overall statistics had been achieved.

Before the $^{46}$V half-life measurement began, we took special precautions to reduce the effect of the 0.12\% $^{42}$Sc contaminant in the $^{46}$V beam exiting from MARS.  These were based on the fact that the $^{42}$Sc ions have a different range in the aluminum absorbers than do the $^{46}$V ions.  To be able to make good use of this fact, we measured the amount of $^{46}$V activity collected in the Mylar tape as a function of the thickness of aluminum degraders as we changed the degrader thickness from 57.2 $\mu$m to 133.4 $\mu$m in increments of 6.4 $\mu$m. This allowed us to determine the thickness of aluminum degraders required to place the collected $^{46}$V activity exactly midway through the tape. The thickness so determined turned out to be very close to the thickness calculated with the code SRIM \cite{zi08} and confirmed the reliability of the code in this application.  We then used SRIM to determine the difference between the range of $^{42}$Sc and that of $^{46}$V, finding the former to be longer by about half the thickness of the Mylar tape.  Thus, by depositing the collected $^{46}$V samples near the back of the Mylar tape we could ensure that $^{42}$Sc passed entirely through the tape.  This virtually eliminated any contribution from $^{42}$Sc to our measured decay spectra.  We will return to this point in Sec.~\ref{ssec:imp}.

\begin{figure}[t]
  \includegraphics[width=\columnwidth]{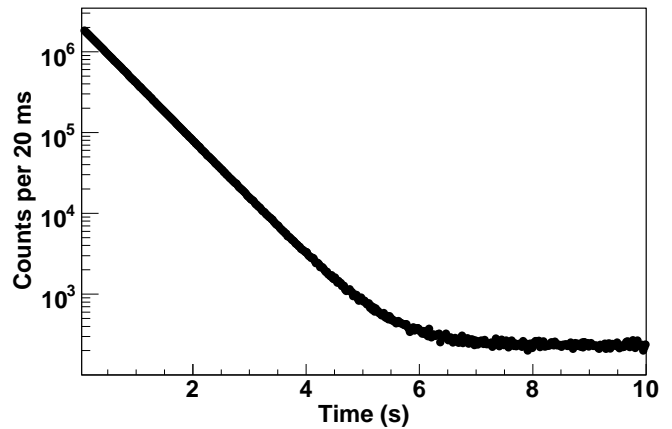}
 \caption{\label{fig2} Measured time-decay spectrum for the total of all data obtained from the $\beta^+$ decay of $^{46}$V.}
\end{figure}

The measurement itself was subdivided into a number of separate runs so that we could thoroughly test for possible systematic effects caused by detection parameters. Run by run, we used different combinations of four dominant dead times (3, 4, 6 and 8 $\mu$s), three discriminator thresholds (150, 200 and 250 mV), and two detector biases (2650 and 2750 V). In all, over 65 million $\beta$ events were recorded from 14422 cycles divided into 16 runs. To illustrate the overall quality of the data, we present in Fig.~\ref{fig2} the total time-decay spectrum obtained by combining all the runs.  In our actual analysis though, we treated the data in each run separately.

As our experiment was aimed at the highest possible precision, several other special precautions were taken as well: 

\begin{enumerate}[(i)]

\item {Our tape-transport system is quite consistent in placing the collected source within $\pm$3 mm of the center of the detector. However, it is a mechanical system and some exceptions occur.  We monitored its performance closely by separately recording the number of nuclei detected in the BC404 scintillator at the exit from MARS during the collection period of each cycle, and the number of positrons detected in the gas counter during the subsequent count period. The ratio of the latter to the former is a sensitive measure of whether the source was seriously misplaced in the proportional counter. We used this ratio as one of the data-selection criteria in our later analysis (see Sec.~\ref{sec:anal}).}

\item {In one run, the decay spectrum was measured with 300 s collect and detect times to probe for unanticipated longer-lived impurities. Apart from $^{46}$V and a low-level constant background, nothing was found.}

\item {A background measurement was made in which all conditions were identical to those of a normal run except that the tape motion was disabled. The background rate at our shielded counting location was $\sim$ 1 count/s, which is 3-4 orders of magnitude lower than the initial count rate for each collected sample and is consistent with the background level observed in normal runs (see Fig.~\ref{fig2}).}

\end{enumerate}

\section{Analysis and Results}
\label{sec:anal}
Before analysis, we pre-selected the data based on two criteria. First, a cycle was rejected if the total number of $\beta$-particles detected by the gas counter was less than 500, indicating that there had been little or no primary beam from the cyclotron during the collection period.  With the rather difficult $^{47}$Ti beam this occurred quite often, so approximately 9\% of the total cycles from all 16 runs were eliminated by this criterion. The second criterion we used to exclude cycles was the measured ratio of detected $\beta$-particles to implanted $^{46}$V ions as observed in the BC404 scintillator.  For each run we restricted the ratio to being between 90 and 100\% of the maximum value obtained for that run; this ensured that the $^{46}$V sample had been centrally located in the detector.  We have already shown this to be a safe range for the ratio \cite{ia10} but, to be doubly sure, we checked it again with this measurement: we found that enlarging the accepted range by a further 10\% had no appreciable effect on the extracted half-life. With the 90-100\% criterion applied to the ratio, another $\sim$8\% of the total cycles were rejected from all 16 runs. The remaining data were then corrected cycle-by-cycle for dead-time losses, based on the method described in Ref.~\cite{ko97}. The final decay spectrum for each run was then obtained from the sum of the dead-time corrected decay spectra for all accepted cycles in that run.

\begin{figure}[t]
\centering
\includegraphics[width=\columnwidth]{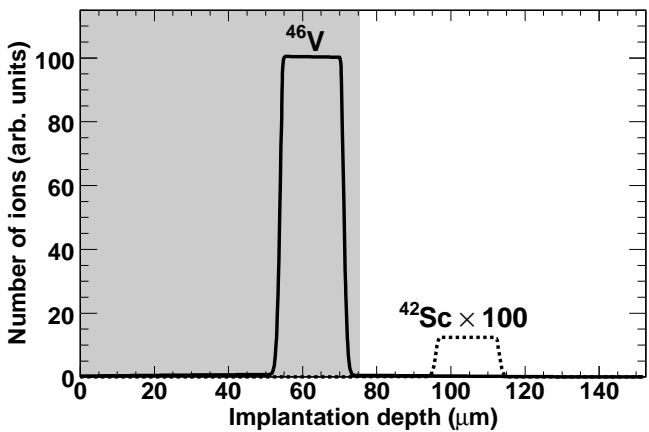}
\caption{\label{fig3} The implantation profiles of $^{46}$V (solid line) and $^{42}$Sc (dashed line) in and beyond the Mylar tape, under the conditions applying to our half-life measurements.  The beams enter from the left. The shaded region corresponds to the actual thickness of our collection tape: all ions within the shaded region are collected in our sample; all others are not.}
\end{figure} 

\subsection{\label{ssec:imp}Sample impurity}

As explained in Sec.~\ref{sec:exp}, $^{42}$Sc is the only impurity in the beam with the potential to affect our half-life measurement.  Other weak impurities seen in the MARS focal plane have half-lives much too long to be of any concern.  We confirmed this conclusion by the decay spectrum taken with 300-s collect and detect times, in which no impurities with intermediate half-lives were detected.  We also explained that we located each $^{46}$V sample at the back of the collection tape to ensure that any $^{42}$Sc present in the beam passed entirely through the tape.

To be more quantitative, we show as the solid line in Fig.~\ref{fig3} a simplified functional form for the implantation depth distribution of $^{46}$V, which is consistent with the results of our scan of $^{46}$V activity versus degrader thickness (see Sec.~\ref{sec:exp}) and with the known momentum spread as set by the momentum slits in MARS ($\Delta p/p=0.88\%$).  We have then reproduced the shape of this empirical depth distribution as a dashed line to represent the distribution of $^{42}$Sc.  The difference in depth between the two was taken from a calculation with the SRIM code~\cite{zi08} and the relative magnitude was set at 0.12\%, the amount we determined from our measurement with the PSSD in the MARS focal plane (see Sec.~\ref{sec:exp}).  Although it is too small to be visible in the figure, our detailed scan of $^{46}$V activity versus degrader thickness showed evidence of a very weak tail, amounting to $\sim$1\% of the total, extending to the left of the depth distribution.  We take this result to be a good gauge of the upper limit for how many $^{42}$Sc nuclei could have been retained in the collection tape.  With this approach we can then conclude that the $^{42}$Sc/$^{46}$V ratio in the collected samples was less than 0.0015$\%$, a result that makes negligible any contribution from $^{42}$Sc to our extracted half-life for $^{46}$V.

\begin{figure}[b]
\centering
\includegraphics[width=\columnwidth]{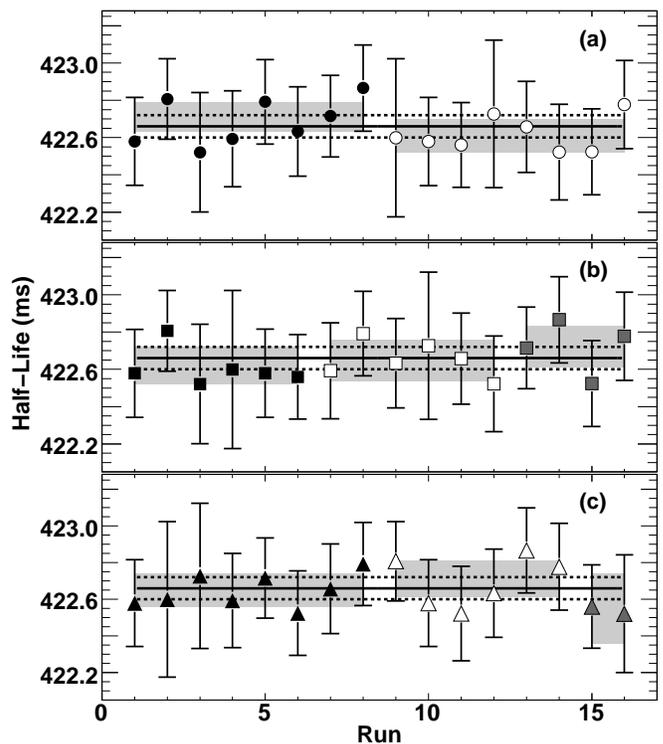}
\caption{\label{fig4} Test for possible systematic bias in the $^{46}$V half-life measurement due to three different detection parameters: (a) two detector biases, 2650\,V/2750\,V, represented by black/open circles; (b) three discriminator settings, 150\,mV/200\,mV/250\,mV, represented by black/open/grey squares; (c) three imposed dead times, 4$\mu$s/6$\mu$s/8$\mu$s, represented by black/open/grey triangles.  Note that the runs have been grouped differently in each part of the figure. In all cases, the grey bands represent the $\pm \sigma$ limits of the average for a given condition. The average value for the half-life is 422.66(6) ms (statistical uncertainty only) with $\chi^2/ndf=3.3/15$. The average value for all the runs appears as the solid line, with dashed lines as uncertainty limits.}
\end{figure}

We also used a second independent method to search for any evidence of $^{42}$Sc activity by examining the recorded time-decay spectra.  We fitted each spectrum from the 16 individual runs with a function including two exponentials, one each for the decays of $^{46}$V and $^{42}$Sc, together with a constant background. In the first fit, we set the initial $^{42}$Sc/$^{46}$V ratio of intensities to the 0.0015$\%$ value just obtained; set the half-life of $^{42}$Sc to its world-average value, 680.72 ms \cite{ha09}; and extracted a half-life for $^{46}$V.  Then we refitted the same 16 spectra with the initial $^{42}$Sc/$^{46}$V ratio as the adjustable parameter.  In this case, the half-life of $^{42}$Sc was again set to its world-average value but the $^{46}$V half-life was fixed at a range of values around the average value obtained from the first fit.  We found that the $^{42}$Sc/$^{46}$V ratio obtained from the fits was very insensitive to the half-life used for $^{46}$V, and that, in all cases, the ratio was less than 0.01$\%$.

In arriving at the final half-life for $^{46}$V and its uncertainty, we have adopted a very conservative range for the $^{42}$Sc/$^{46}$V ratio, taking the value to be 0.006(6)\%.

\subsection{\label{ssec:tdec}Time-decay analysis} 

\begin{figure}[t]
\centering
\includegraphics[width=\columnwidth]{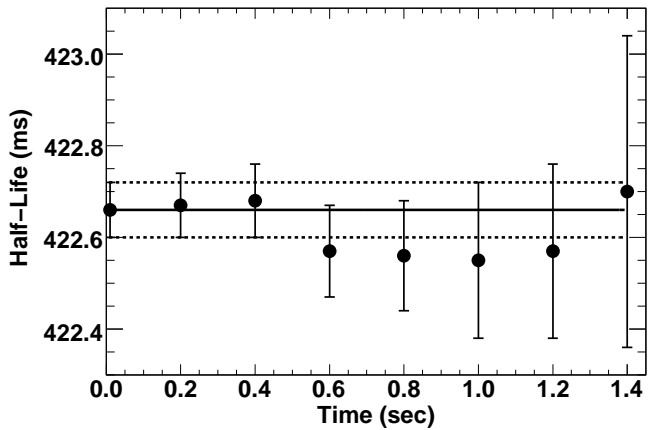}
\caption{\label{fig5} Test for possible systematic bias in the $^{46}$V half-life measurement caused by short-lived impurities or by unexpected rate-dependent counting losses. Each point is the result of a separate fit to the data; the abscissa for each point represents the time period at the beginning of the counting cycle for which the data were omitted from that particular fit.  The solid and dashed lines correspond to the average half-life value and uncertainties for the full data set (see Fig.\,\ref{fig4}).}
\end{figure} 

We fitted the data from each of the 16 runs separately, incorporating three components: $^{46}$V, $^{42}$Sc and a constant background.  The half-life of 
$^{42}$Sc was fixed at its known value of 680.72 ms \cite{ha09} and, as explained in the last section, the initial activity of $^{42}$Sc relative to $^{46}$V was set at 0.006$\%$.  Since each run was obtained with a different combination of detection settings, we could use the individually fitted half-lives of $^{46}$V to test for any systematic dependence on those settings.  As displayed in Fig.~\ref{fig4}, the half-life results showed no systematic dependence on detector bias voltage, discriminator threshold setting, or the dominant dead time we imposed in the electronics, the average half-life yielding a remarkably low value for the normalized $\chi^2$ of 0.2.  With this degree of consistency in the data, we can be confident that any systematic dependence on detection parameters must be negligible with respect to our quoted statistical uncertainty.

Our final systematic check was to test for unanticipated short-lived impurities in the decay spectrum or for any other evidence showing count-rate dependence of the half-life. In each run we removed data from the first 0.2 s of the counting period and fitted the remaining data; then we removed an additional 0.2 s of data and refitted again. This procedure was continued until 1.4 s of data -- 3.3 half-lives of $^{46}$V -- had been removed.  The average half-life from all 16 runs was then obtained for each set of truncated data.  The result is given in Fig.~\ref{fig5} where it can be seen that the half-lives so obtained were consistent with one another within statistical uncertainties.

\begin{figure}[t]
\centering
\includegraphics[width=\columnwidth]{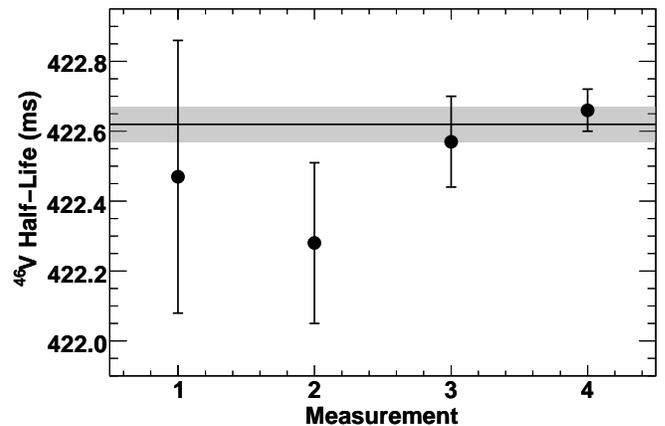}
\caption{\label{fig6} The present measurement is compared with all the published measurements of $^{46}$V half-lives with quoted uncertainties that are within a factor of 10 of ours. The results are presented in chronological order from left to right. The $\pm \sigma$ limits on the overall average value of 422.62(5) ms appear as the grey band.}
\end{figure} 

With these possible systematic effects eliminated as significant factors, our final result for the half-life of $^{46}$V is 422.66(6) ms. As shown in Table \ref{ebudget}, the quoted uncertainty is dominated by counting statistics since the only other identifiable contribution -- from a possible $^{42}$Sc impurity -- is substantially smaller even when very generous limits have been applied (see Sec.~\ref{ssec:imp}).

\begin{table}[b] 
\caption{\label{ebudget} Error budget for the $^{46}$V half-life measurement.}
\begin{ruledtabular}
\begin{tabular}{lc}
\multicolumn{1}{l}{Source}&
\multicolumn{1}{c}{Uncertainty (ms)} \\
\hline
statistics                      & 0.06 \\
sample impurity ($^{42}$Sc)     & 0.02 \\
Total                           & 0.06 \\ 
\\
$^{46}$V half-life result (ms)  & 422.66(6) \\
\end{tabular}
\end{ruledtabular}
\end{table}

The quoted precision of our result is 0.014$\%$.  There have also been three previous measurements claiming sub 0.1\% precision; they yielded half-life values of 422.47(39) ms~\cite{al77}, 422.28(23) ms~\cite{ba77}, and 422.57(13) ms~\cite{ko97}, all less precise than ours.  The first two of these measurements depended on plastic scintillators to detect $\beta$ particles from the decay, while the third used a very similar technique to ours, with a proportional gas counter used for $\beta$ detection.  Only the latter experiment had assured purity of the decaying sample, having employed an on-line isotope separator to produce it.  These three previous results are compared with our new value in Fig.~\ref{fig6}.  All four values are statistically consistent with one another, with only the result from Ref.~\cite{ba77} lying slightly outside of one standard deviation from the overall average, which is 422.62(5) ms.  The normalized $\chi^2$ for the average is a satisfactory 1.0. 

\section{Conclusions}
We have measured a new value for the half-life of the superallowed $\beta^{+}$ emitter $^{46}$V.  Our result, 422.66(6) ms, is a factor of two more precise than the best previous result, with which it is completely consistent.  If our half-life value is combined with previous measurements of the same quantity, it yields a new world average of 422.62(5) ms.  The other properties of this decay -- its $Q_{EC}$ value, branching ratio and the calculated corrections for radiative and isospin-symmetry breaking effects -- have been tabulated recently in Ref.~\cite{ha09}.  Combining the new world-average half-life and the new $Q_{EC}$ value from Ref.~\cite{Er11} with these tabulated values we obtain a corrected $\mathcal{F}t$ value for the $^{46}$V superallowed transition of 3074.1(26)\,s, where the uncertainty is dominated by the uncertainties applied to the theoretical correction terms. This result is entirely consistent with the average $\overline{\mathcal{F}t}$ value of 3072.08(79) s established in the most recent survey~\cite{ha09} of world data from 13 well-known superallowed transitions.

Evidently, the important experimental components of the $^{46}$V superallowed transition -- its half-life and $Q_{EC}$ value -- have now been satisfactorily confirmed and improved.  The $\mathcal{F}t$ value for this transition is certainly not anomalous at the 0.08\% level of precision currently quoted on that quantity.

\begin{acknowledgments}
We thank Dr. D.P. May and Dr. G.J. Kim for developing the challenging $^{47}$Ti beam used in our measurement.
This work was supported by the U.S. Department of Energy under Grant No.\,DE-FG03-93ER40773 and by the Robert
A. Welch Foundation under Grant No.\,A-1397. 

\end{acknowledgments}

\end{document}